\documentclass[%
 reprint,
 groupedaddress,
 amsmath,amssymb,
 aps,
 prd,
]{revtex4-1}

\usepackage{graphicx}
\usepackage{hyperref}
\usepackage[utf8]{inputenc}

\newcommand{\mnras}{MNRAS}
\newcommand{\jcap}{JCAP}
\newcommand{\apjs}{Astrophys. J. S.}

\begin{document}

\title{Cosmological particle-in-cell simulations with ultralight axion dark matter}

\author{Jan Veltmaat}
\author{Jens C. Niemeyer}
\affiliation{%
 Institut f\"ur Astrophysik\\
 Universit\"at G\"ottingen
}%

\date{\today}

\begin{abstract}
We study cosmological structure formation with ultralight axion dark matter, or ``fuzzy dark matter'' (FDM), using a particle-mesh scheme to account for the quantum pressure arising in the Madelung formulation of the Schrödinger-Poisson equations. Subpercent-level energy conservation and correct linear behavior are demonstrated.  Whereas the code gives rise to the same core-halo profiles as direct simulations of the Schrödinger equation, it does not reproduce the detailed interference patterns. In cosmological simulations with FDM initial conditions, we find a maximum relative difference of
O($10\%$) in the power spectrum near the quantum Jeans length compared
to using a standard N-body code with identical initial conditions. This shows that the effect of quantum pressure during nonlinear structure formation cannot be neglected for precision constraints on a dark matter component consisting of ultralight axions.
\end{abstract}

\pacs{Valid PACS appear here}
\maketitle

\section{\label{sec:intro} Introduction}

Ultralight axions (ULAs) are candidate particles for dark matter motivated by string theory compactifications \cite{Arvanitaki2010} that behave as ``fuzzy dark matter'' (FDM) \cite{Hu2000}. 
In these models, the gravitational growth of perturbations is suppressed below the ``quantum Jeans length'' corresponding to the de Broglie wavelength of particles with virial velocities. This effect has also been studied in the contexts of scalar field \cite{Suarez2011,Robles2012,Rindler-Daller2014,Lora2014,Martinez-Medina2015}, Bose-Einstein condensate \cite{Chavanis2011,Guzman2015,Lee2016}, or wave dark matter \cite{Chen2016}.
For sufficiently light particles ($m \lesssim 10^{-22}$ eV), the Jeans length is of the order of several kpc or larger, allowing mass-dependent constraints on the relative fraction of FDM to CDM from observations of galactic density profiles, halo statistics, large-scale structure, and the CMB \cite{Marsh2015a}. Moreover, mixed dark matter models consisting of FDM and CDM have been proposed to solve the small scale problems of pure CDM scenarios \cite{Hu2000, Peebles2000, Marsh2014} including the missing satellites \cite{Moore1999,Klypin1999} and cusp-core problems \cite{Gilmore2007}. ULAs share the bosonic properties and the vacuum realignment production mechanism with the QCD-axion. The mass of the latter is, however,
constrained by black hole superradiance to $m > 10^{-11}$ eV \cite{Arvanitaki2015}, pushing the quantum Jeans length to scales too small to be observable in cosmological phenomena.

The currently strongest constraints on the scalar field mass have been obtained from the luminosity function of high-$z$ galaxies and measurements of the optical depth to reionization \cite{Bozek2015,Schive2016,Sarkar2016}. The resulting minimal mass of $m \ge 10^{-22}$ still produces sufficiently large halo cores to agree with observations \cite{Marsh2014,Schive2014a,Marsh2015c,Calabrese2016,Chen2016}. Thus, FDM currently appears not to suffer from the catch-22 problem as warm dark matter does \cite{Maccio2012}.

In the nonrelativistic limit, the dynamics of FDM is governed by the coupled Schrödinger-Poisson (SP) equations whose ground state solutions, equivalent to the well-known (nonrelativistic) boson stars, have been studied extensively in spherical or cylindrical symmetry \cite{Ruffini,Guzman2004,Bernal2006a,Bernal2006,Liebling2012,Gonzales2011,Guzman2015}.
Only recently, simulations of the comoving SP equations with cosmological initial conditions have become feasible \cite{Woo2009,Schive2014}. They show the formation of dark matter halos consisting of solitonic cores with ground state (boson star) density profiles, embedded in an NFW-like outer halo. Three-dimensional simulations of merging solitonic cores have confirmed this result \cite{Schive2014a, Schwabe2016}. 

In order to resolve oscillatory patterns carrying momentum information on scales of $(mv)^{-1}$, cosmological simulations based on the discretized SP equations require computational resources that currently only allow small box sizes $\sim 1$ Mpc$^3$, making predictions based on halo statistics practically impossible.
On the other hand, like in warm dark matter scenarios, some constraints rely predominantly on delayed structure formation and the absence of low-mass halos rather than the existence of halo cores. These are captured to some extent by the linear transfer function, whereas dynamical effects during nonlinear gravitational collapse produce only higher-order corrections. Following this argument, standard N-body simulations with FDM initial conditions have been used to derive constraints on scalar field masses from reionization and damped Lyman-$\alpha$ observations \cite{Schive2016,Sarkar2016}. However, the accuracy of this approach has not been tested so far for FDM.

In this work, we present a particle-mesh scheme including an additional force term to account for the quantum pressure gradient in the Madelung formulation of the Schrödinger equation. It is intended to provide a coarse-grained description of the SP dynamics instead of a fully accurate alternative representation by keeping track of the exact energy budget without resolving the fine-scale structure. It aims to model the effects of quantum pressure on scales where differences between the dynamics of CDM and FDM are moderate. The method can be understood as an intermediate approach between grid-based Schrödinger solvers and N-body codes. Because the quantum pressure depends on the second derivative of the density field it is extremely sensitive to small scale fluctuations. To mitigate this effect, we choose a discretization which inherently conserves energy. An alternative approach to simulate the SP equations in the Madelung form with a particle-based method can be found in \cite{Mocz2015}, see also the discussion in \cite{Marsh2015b}.

The paper is organized as follows. In \autoref{sec:methods}, the employed numerical method and parameters are described. The method is then tested in noncosmological test problems. In \autoref{sec:results} the results of our cosmological runs are presented. The results are summarized and discussed in \autoref{sec:conclusion}.

\section{Numerical methods}
\label{sec:methods}

\subsection{Particle-mesh implementation of quantum pressure}
\label{sec:part-mesh-impl}

Nonrelativistic FDM obeys the Schrödinger-Poisson equations,
\begin{align}
 i\hbar\dot\Psi = -\frac{\hbar^2}{2m} \nabla^2 \Psi + V m \Psi
\end{align}
and
\begin{align}
 \nabla^2 V = 4\pi G m |\Psi|^2\,\,.
\end{align}
The Schrödinger equation can be recast using the expressions $\Psi = \sqrt{\frac{\rho (\mathbf x ,t)}{m}} \exp(iS(\mathbf x,t)/\hbar)$ and $\mathbf v = m^{-1} \nabla S$. Separating real and imaginary parts yields
\begin{align}
\label{eq:massconservation}
\dot \rho + \nabla (\rho \mathbf v) = 0
\end{align}
and
\begin{align}
\label{eq:euler}
\dot{\mathbf v} + (\mathbf v \cdot \nabla) \mathbf v = - \nabla (Q+V)\,\,,
\end{align}
where the so-called quantum pressure is given  by
\begin{align}
\label{eq:quantumpressure}
Q = -\frac{\hbar^2}{2m^2} \frac{\nabla^2 \sqrt{\rho}}{\sqrt{\rho}}\,\,.
\end{align}
This representation resembles the Euler equations of fluid dynamics. It is known as the Madelung transformation of the Schrödinger equation \cite{Madelung1927}.
Both formulations are equivalent in regions of nonvanishing density, i.e., outside of interference nodes where $Q$ is ill defined. Consequently, interference phenomena are still expected to occur in high-resolution solutions of the Madelung equations, but an accurate representation of nodes can only be achieved in the Schr\"odinger formulation.

For our simulations, we modified the particle-mesh scheme of the cosmology code \textsc{Nyx} \cite{Almgren2013} in order to account for the quantum pressure $Q$.  In analogy to smoothed-particle hydrodynamics (SPH) \cite{Monaghan2005}, we derive an equation of motion for the particles from a Lagrangian which ensures inherent energy conservation. To define an adequate Lagrangian, an expression for the kinetic energy of the scalar field under the Madelung transformation is needed. It is given by
\begin{align}
\label{madkinen}
 K &= \int \frac{\hbar^2}{2m} |\nabla \Psi|^2 \text{d}^3\mathbf x\nonumber\\ 
&=  \int \frac{\rho}{2} \mathbf{v}^2 \text{d}^3 \mathbf x  + \int \frac{\hbar^2}{2m^2} (\nabla \sqrt{\rho})^2 \text{d}^3 \mathbf x =: K_v + K_\rho
\end{align}
Here, $K_v$ will be numerically represented by the kinetic energy of the the particles. $K_\rho$ is the energy carried by the density gradient. It can also be understood as the potential energy corresponding to the quantum pressure. Neglecting gravity, the Lagrangian therefore reads
\begin{align}
L= T-V = K_v - K_\rho\,\,.
\end{align}

Writing the Lagrangian in terms of the positions $\mathbf{q}_i$, velocities $\mathbf{\dot{q}}_i$ and masses $m_i$ of the particles with indices $i$, we find for $K_v$:
\begin{align}
\label{eq:diskkinen}
K_v = \sum_i \frac{1}{2} m_i \mathbf{\dot{q}}_i^2\,\,.
\end{align}
For $K_\rho$, we use the expression for the density at a given point $\mathbf{x}$  used in SPH codes \cite{Monaghan1992},
\begin{align}
\label{eq:rhodis}
 \rho(\mathbf{x}) = \sum_i m_i W(|\mathbf{x}-\mathbf{q}_i|,h)\,\,,
\end{align}
where $W$ is the smoothing kernel with a given smoothing length $h$. Thus,
\begin{align}
 K_\rho = \int \frac{\hbar^2}{2m^2} \left(\nabla \sqrt{\sum_i m_i W(|\mathbf{x}-\mathbf{q}_i|,h)}\right)^2 \text{d}^3\mathbf{x}\,\,.
\end{align}

The gradient is replaced by finite differences and integrals are evaluated on the rectangular grid that is already used in the particle-mesh scheme for gravity in \textsc{Nyx},
\begin{align}
\label{eq:diskpoten}
 K_\rho \approx &\frac{\hbar^2}{2m^2} \sum_{j,k,l} \left( \nabla \sqrt{ \rho(\mathbf{x})} \big|_{\mathbf{x}=\mathbf{x}_{jkl}} \right)^2 (\Delta x)^3\nonumber\\
 \approx &\frac{\hbar^2}{2m^2} \frac{\Delta x}{4} \sum_{j,k,l} \left(\sqrt{\rho_{j+1,k,l}} - \sqrt{\rho_{j-1,k,l}} \right)^2\nonumber\\ 
&+ \left(\sqrt{\rho_{j,k+1,l}} - \sqrt{\rho_{j,k-1,l}} \right)^2\nonumber \\ &+ \left(\sqrt{\rho_{j,k,l+1}} - \sqrt{\rho_{j,k,l-1}} \right)^2 \,\,.
\end{align}
Here, $\mathbf{x}_{jkl}$ denote the grid points with a separation of $\Delta x$ and $\rho_{j,k,l} \equiv \rho(\mathbf{x}_{j,k,l})$. Using the second Lagrange equation,
\begin{align}
 \frac{\text{d}}{\text{d}t} \frac{\partial L}{\partial \mathbf{\dot{q}}_i} - \frac{\partial L}{\partial \mathbf{q}_i} = 0\,\,,
\end{align}
and noting that 
\begin{align}
 \frac{\partial \rho(\mathbf x) }{\partial \mathbf{q}_i} &= \frac{\partial}{\partial \mathbf{q}_i} \sum_i m_i W(|\mathbf x - \mathbf{q}_i|,h) \nonumber \\ &= - m_i \nabla W(r,h)\big|_{\mathbf x - \mathbf{q}_i}\,\,,
\end{align}
yields the following equation of motion:
\begin{align}
\label{eq:equationofmotion}
 m_i \mathbf{\ddot{q}}_i
  =  \frac{\hbar^2}{2m^2} (\Delta x)^3 \sum_{j,k,l} \frac{\Delta_n \sqrt{\rho}}{\sqrt{\rho}} m_i \nabla W(r,h) \big|_{\mathbf{x}_{j,k,l} - \mathbf{q}_i}\,\,.
\end{align}
$\Delta_n$ is the numerical Laplace operator (seven-point stencil). One can therefore recognize the quantum pressure evaluated by finite differences on the right hand side of \autoref{eq:equationofmotion} . By construction, the discretized equation of motion conserves the discretized versions of the energy expressions as given by \autoref{eq:diskkinen} and \autoref{eq:diskpoten}. The additional force term was incorporated into the existing leapfrog scheme for gravitational acceleration in \textsc{Nyx}. The principal steps for its computation are to compute (i) the density on the grid using equation (\autoref{eq:rhodis}) (ii) the quantum pressure on the grid using the seven-point stencil and \autoref{eq:quantumpressure}, and (iii) the acceleration on each particle by
convolving the pressure with the gradient of the smoothing kernel according to \autoref{eq:equationofmotion}.

The same smoothing kernel is used for the density interpolation in the gravity solver, so that the density field in \textsc{Nyx} is always computed by \autoref{eq:rhodis}.
The additional acceleration leads to a modified condition for the time step $\Delta t$ in order to enforce that particles must not move farther than $\Delta x$ in a single time step \cite{Almgren2013}.

We choose the following smoothing kernel \cite{Monaghan1992}:
\begin{align}
 W(r,h) = \frac{8}{\pi h^3}
 \begin{cases}
  1 - 6 \left(\frac{r}{h}\right)^2 + 6 \left(\frac{r}{h} \right)^3 &\text{if}~  0 \leq r/h \leq \frac{1}{2}\\
  2 \left( 1 - \frac{r}{h} \right)^3 & \text{if} ~ \frac{1}{2} < r/h \leq 1 \\
  0 & \text{elsewhere}
 \end{cases}
\end{align}
$h \approx 4\Delta x$ was empirically found to optimize energy conservation and computational efficiency and therefore chosen for the simulations of solitonic core collisions. In the cosmological simulations, we used $h = 3.5 \Delta x$ since kernel radii with an uneven number of cells give rise to a less noisy density field after initialization with the Zel'dovich approximation. This is especially important for FDM initial conditions featuring a steep cut-off in the initial power spectrum. We assume a scalar field mass of $m = 2.5 \times 10^{-22}$ eV throughout this paper.

\subsection{Tests with solitonic core collisions}
\label{sec:tests-with-solitonic}

In \cite{Schwabe2016}, three-dimensional simulations of the SP equations were used for a detailed parameter study of colliding and merging solitonic halo cores. As shown in \cite{Schive2014}, the density structure of cores in FDM halos is identical to Newtonian boson stars. Here, we will use the results of  \cite{Schwabe2016} to compare the particle-mesh code described above to numerical solutions of the discretized SP equations in a nonsymmetric, dynamical setup on a static background. For details about the Schrödinger solver, see \cite{Schwabe2016}.
As long as no additional scale is introduced (like the Hubble scale in \autoref{sec:results}), the system obeys a scale symmetry with respect to a parameter $\lambda$ \cite{Guzman2004}:
\begin{align}
\label{eq:Skalierung1}
 \{ t , \mathbf{x} , V, \Psi\} \rightarrow \{\lambda^{-2} t , \lambda^{-1} \mathbf{x} , \lambda^2 V, \lambda^2 \Psi \} \nonumber \\
 \{ \rho , M , K, W_g\} \rightarrow \{\lambda^{4} \rho , \lambda M , \lambda^3 K, \lambda^3 W_g \}\,\,.
\end{align}
Here, $M$ denotes the total mass and $W_g$ the gravitational potential energy. Thus, although we present our results for a concrete set of physical parameters, they can be rescaled according to \autoref{eq:Skalierung1}.

\begin{figure}
\centering
 \includegraphics{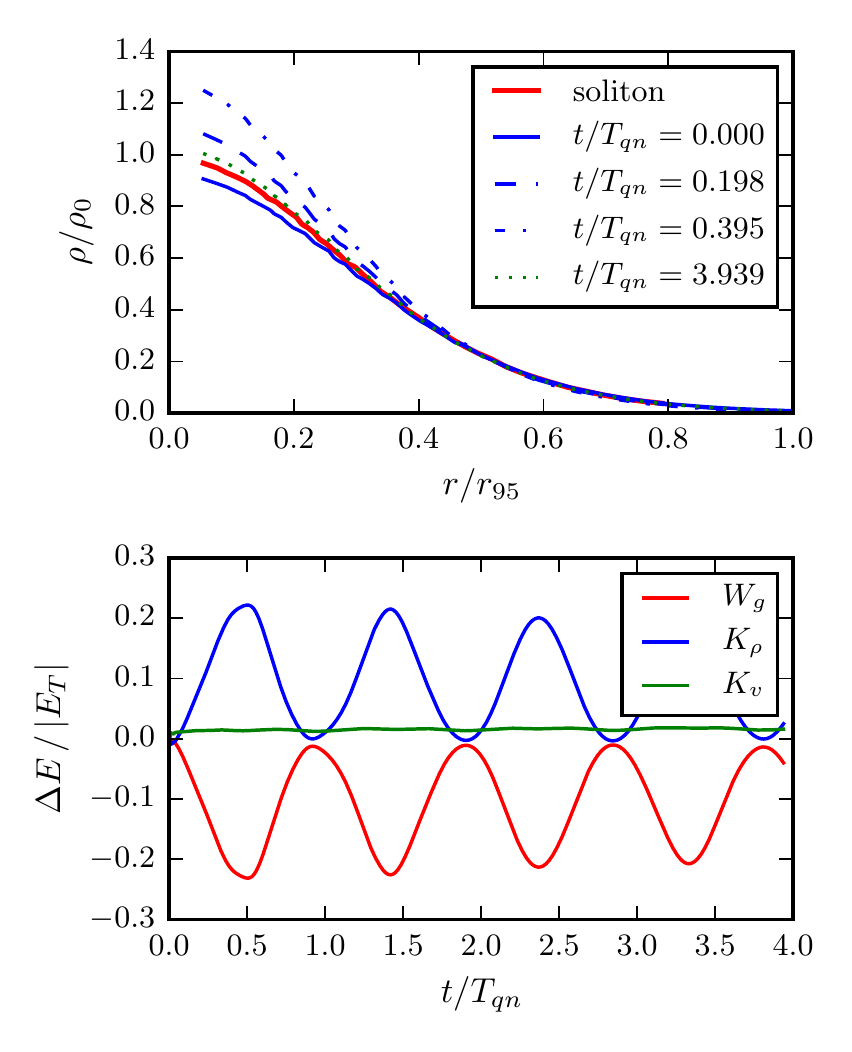}
 \caption{\textit{top:} Evolution of the radial density profile (blue and green) normalized to the central density of the theoretical soliton profile (red) \textit{bottom:} Difference of the three energy terms from their starting values normalized to the total energy of the system}
 \label{fig:plot1}
\end{figure}

For the simulation shown in \autoref{fig:plot1}, we initialized a density profile which is close to the spherically symmetric ground state solution of the SP equations as given in \cite{Guzman2004} which we will refer to as (solitonic) cores. We use $N=10^6$ particles and a grid resolution of $18.6 \Delta x = r_{95}$ where $r_{95}$ is the radius that encloses $95\%$ of the mass of the core. As expected, the code conserves total energy up to less than $10^{-2}$.  Owing to small numerical deviations from the exact solution, the density profile oscillates stably around the ground state. Considering the different energy terms, this behavior corresponds to a periodic exchange between the gradient energy $K_\rho$ and the gravitational potential energy $W_g$. The characteristic (quasinormal) period $T_{qn}$ seen in our solutions is very close to that given by \cite{Guzman2004}.

\begin{figure}
\centering
 \includegraphics{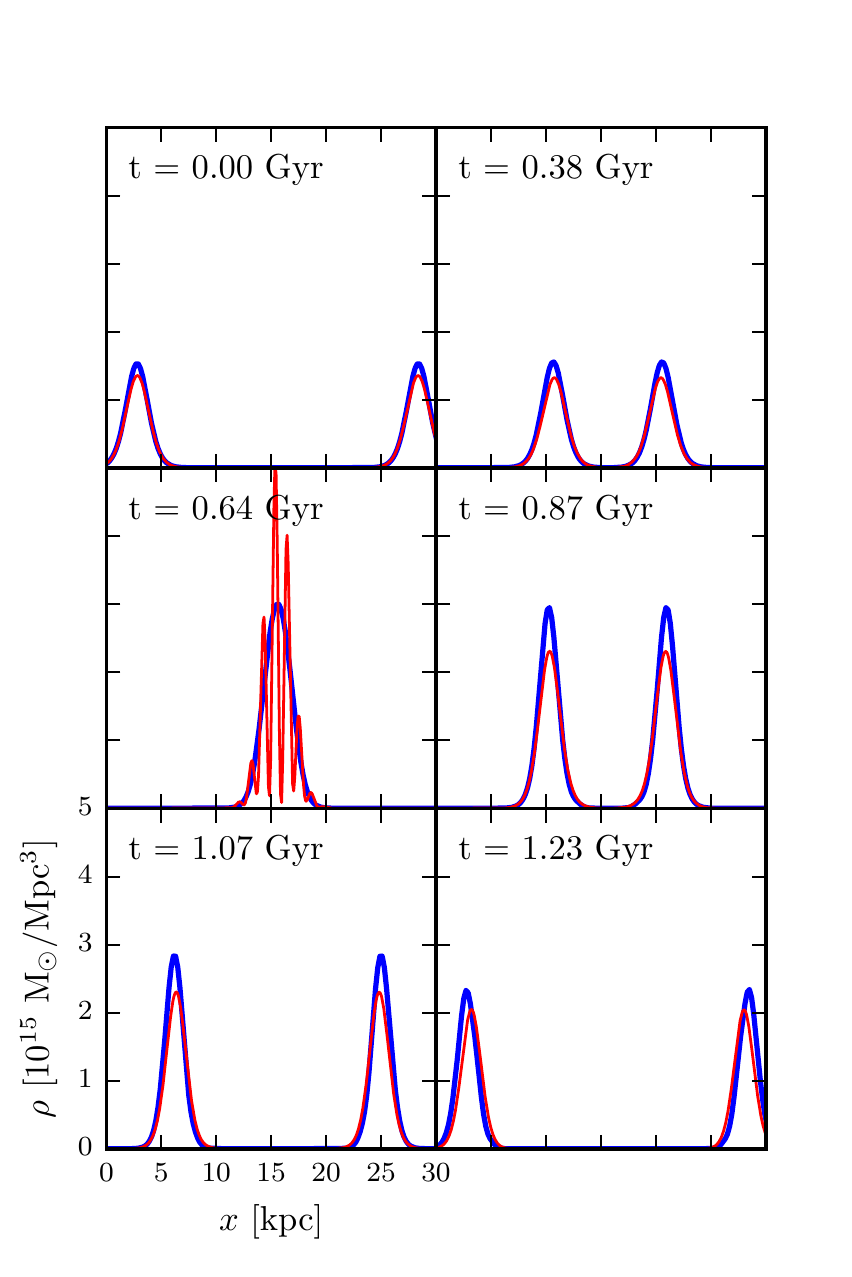}
 \caption{Evolution of the density profile along the symmetry axis for our code (blue) and the Schrödinger solver (red). The cores have an initial relative velocity of 40 km/s.}
 \label{fig:plot2}
\end{figure}

\autoref{fig:plot2} shows a head-on collision of two cores, using the same number of particles per core and spatial resolution as above. Due to their initial relative velocity the total energy is positive. According to \cite{Bernal2006}, one expects the cores to exhibit solitonic behavior, i.e. to pass through each other without significantly altering their density profiles. While overlapping, the cores are expected to form an interference pattern. In figure \autoref{fig:plot2}, our results are compared to those obtained from the grid-based Schrödinger solver used in \cite{Schwabe2016}. As expected from the discussion in \autoref{sec:part-mesh-impl} above, the coarse-grained nature of the particle-mesh method does not capture the interference pattern but nevertheless reproduces the solution of the Schrödinger solver before and after the cores overlap.

\begin{figure}
\centering
 \includegraphics[width=\columnwidth]{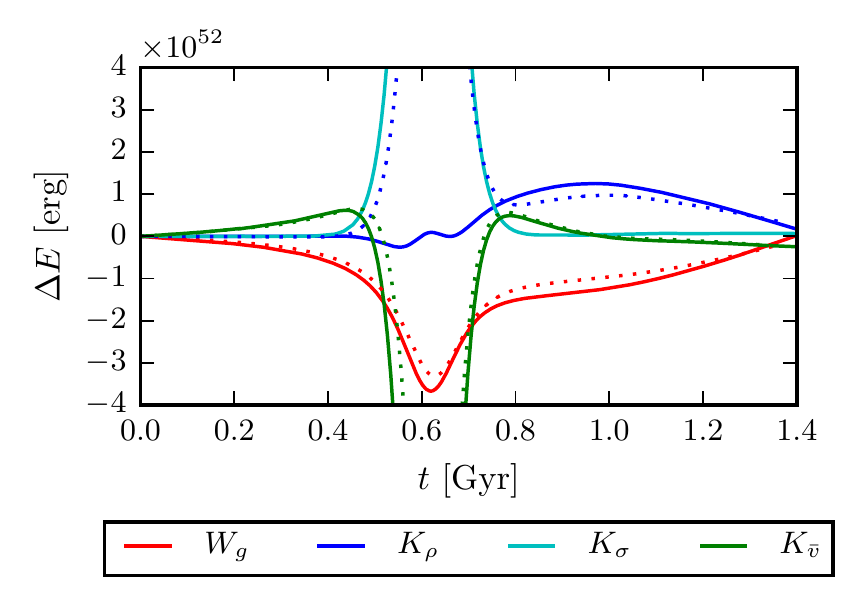}
 \caption{Difference of the energy terms to their initial values in the simulation shown in figure \autoref{fig:plot2} for our code (solid lines) and the Schrödinger solver (dotted lines).}
 \label{fig:plot3}
\end{figure}

In \autoref{fig:plot3}. we plot the energy terms for both codes where, in the case of the particle-mesh scheme, we split $K_v$ into the part resulting from the grid-scale bulk motion of the fluid, defined by
\begin{align}
 \mathbf{v}(\mathbf{x}) = \frac{\sum_i \mathbf{\dot q}_i m_i W(|\mathbf{x}-\mathbf{q}_i|,h)}{\rho(\mathbf{x})}\,\,,
\end{align}
and the remaining part corresponding to the particle velocity dispersion. The most distinguishable feature is that during overlapping phase, energy is stored in two different ways. While in the Schrödinger solution gradient energy increases as a result of the formation of an interference pattern, in the particle-mesh scheme the exact same amount of energy is stored in particle velocity dispersion. In other words, on scales where the true density profile is unresolved, particles are largely unaffected by quantum pressure and therefore stream freely. This is in obvious contrast with solutions of \autoref{eq:euler} where $\mathbf{v}$ is single-valued at each point. Our numerical scheme can therefore be interpreted as an approximate, coarse-grained representation of the SP equations whose validity on the scales of interest must be verified by numerical tests. Our results show that energy conservation is crucial for establishing the correct behavior on resolved scales.

\begin{figure}
\centering
 \includegraphics[width=\columnwidth]{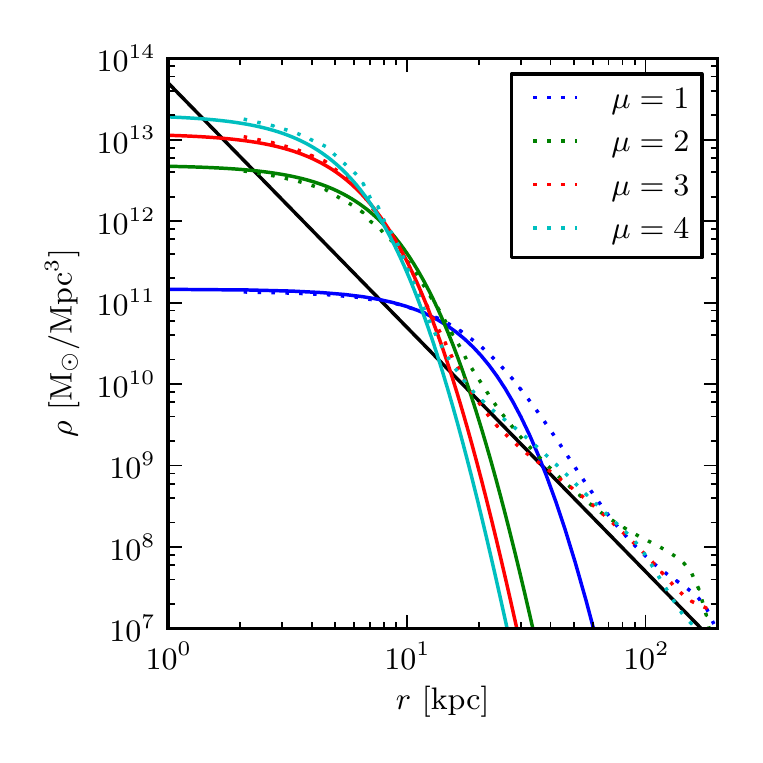}
 \caption{Radial profiles of the resulting haloes in simulations of two merging cores (dotted line). The solid lines are fitted theoretical core profiles. The black line shows a $r^{-3}$ law as in the outer regions of a NFW profile.  The initial set-up and resolution is equal to the one in figure \autoref{fig:plot2} but with an initial velocity of $0.1$ km/s in opposite directions perpendicular to the symmetry axis instead of a velocity towards each other. The runs have different mass ratios $\mu$ between the two cores where the lighter one always has the same mass as in figure \autoref{fig:plot2}.}
 \label{fig:radialprofile}
\end{figure}

Further evidence for this interpretation is provided by \autoref{fig:radialprofile}. It shows the radial density profiles of the final state in simulations of two merging cores with initial orbital angular momentum. In agreement with the results of SP simulations \cite{Schive2014a,Schwabe2016} , they consist of a central solitonic core and an outer halo with NFW-like profile. As expected, we do not observe the granular structure surrounding the core seen in the SP simulations. As in \cite{Schwabe2016}, the mass of the central core is independent of angular momentum.  

\section{Cosmological simulations}
\label{sec:results}

In order to isolate the dynamical impact of quantum pressure, we compare cosmological simulations using the particle-mesh scheme (``FDM simulations'') with standard N-body simulations (apart from the unusual smoothing kernel) using identical initial conditions (``CDM simulations''). In comoving coordinates, the quantum pressure \autoref{eq:quantumpressure} is given by
\begin{align}
\label{eq:comquantumpressure}
Q = -\frac{\hbar^2}{2m^2} \frac{1}{a^2}\frac{\nabla^2 \sqrt{\rho}}{\sqrt{\rho}}
\end{align}
Initial conditions for particles in the Zel'dovich approximation are generated with the \textsc{Music} code \cite{Hahn2011}. We set $H_0 = 70$ km s$^{-1}$ Mpc$^{-1}$, $\Omega_\Lambda = 0.75$ and $\Omega_{\rm m} = \Omega_{\rm FDM} = 0.25$ for our test calculations. In the presentation of the results, we use the internal unit system of the \textsc{Nyx} code and choose Mpc for the unit of length instead of \mbox{Mpc h$^{-1}$}.

\begin{figure}
\centering
 \includegraphics{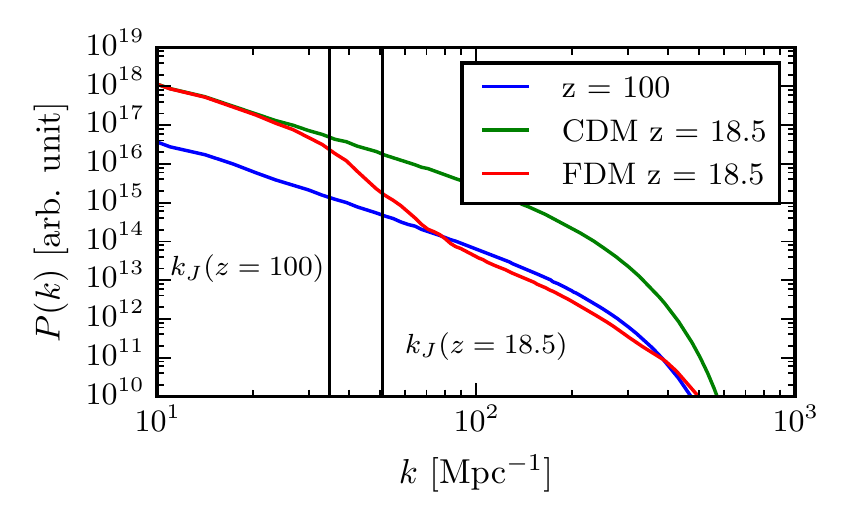}
 \caption{Power spectra in simulations with an initial CDM power spectrum at $z=100$ and the Jeans scales $k_J$ at $z=100$ and $z=18.5$.}
 \label{fig:plot5}
\end{figure}

As a first test, we initialize the simulations with a standard CDM transfer function as parameterized in \cite{Eisenstein1998}. By keeping small scales in the inital power spectrum unsuppressed, the effect of quantum pressure on the evolution of perturbations can be observed more clearly. The comoving box size is $L=2$ Mpc resolved by $N_g = 512^3$ cells. \autoref{fig:plot5} compares the power spectra at $z=100$ and $z=18.5$ where the perturbations are only mildly nonlinear. While the power on large scales is equal in both cases, on smaller scales the power spectrum remains at approximately the initial level in the FDM simulation. \autoref{fig:plot5}  also shows the quantum Jeans scale \cite{Hu2000},
\begin{align}
\label{eq:jeansscale}
 k_J = 2\pi^{1/4} a^{1/4} \hbar^{-1/2} (G\rho)^{1/4}m^{1/2}\,\,,
\end{align}
at the corresponding redshifts. As expected, the region between suppressed and unsuppressed growth of perturbations coincides with the range of $k_J$ between $z=100$ and $z=18.5$ as a result of its dependence on the scale factor.

\begin{figure*}
\centering
 \includegraphics{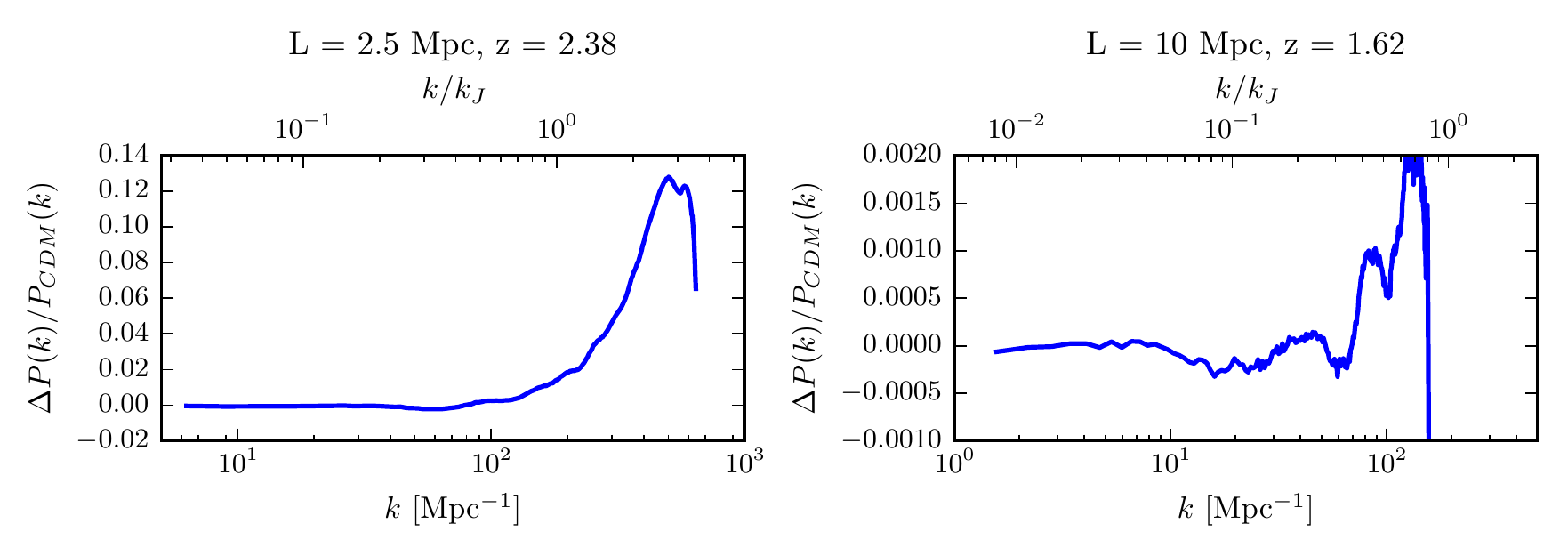}
 \caption{Relative difference in the power spectra of simulations with and without quantum pressure  $(P_{FDM}(k)-P_{CDM}(k))/P_{CDM}(k)$. The scale is also shown relative to the Jeans scale $k_J$ corresponding to the typical density of filaments $\rho = 10^{12}$ M$_\odot$/Mpc$^3$}
 \label{fig:plot6}
\end{figure*}

Accounting for the suppression of small scale power in linear structure growth since the beginning of structure formation can be expressed by a transfer function $T(k,z)$ modifying the CDM power spectrum,
\begin{align}
 P_{FDM}(k,z) = T^2(k,z)P_{CDM}(k,z)\,\,,
\end{align}
which is approximately given by the redshift-independent analytical expression \cite{Hu2000}
\begin{align}
\label{eq:transferhu}
 T(k) = \frac{\cos x^3}{1 + x^8}
\end{align}
where $x = 1.61\, m^{1/18} k/k_{Jeq}$. It corresponds to a sharp cut-off at the Jeans scale at matter-radiation equality $k_{Jeq}$. Due to the qualitative difference in the linear power spectra of FDM and CDM already at the beginning of the simulations, the results in \autoref{fig:plot5} clearly cannot be taken as a realistic simulation of FDM.

As a second, more realistic test of the effects of FDM quantum pressure, we used \autoref{eq:transferhu} to initialize the density fields. The pressureless (CDM) simulations are therefore equivalent to the N-body simulations presented in \cite{Schive2016,Sarkar2016}. In order to investigate the scale dependence of the effects of quantum pressure at fixed axion mass, we used two simulation boxes with $L=2.5$ Mpc and $L=10$ Mpc resolved by $N_G=512^3$ cells. Because $k_J(z=100) > k_{Jeq}$, power at the Jeans scale at $z=100$ is strongly suppressed. Thus, although we resolve the Jeans scale, the influence of quantum pressure on linear structure growth is not as visible as in \autoref{fig:plot5} because there is not as much structure on the relevant scales at initialization. Instead, we must look for differences in nonlinear structure growth. Note, however, that the employed resolution is not sufficient to resolve the solitonic cores in dark matter halos as observed in \cite{Schive2014,Schive2014a}.

\begin{figure*}
\centering
 \includegraphics{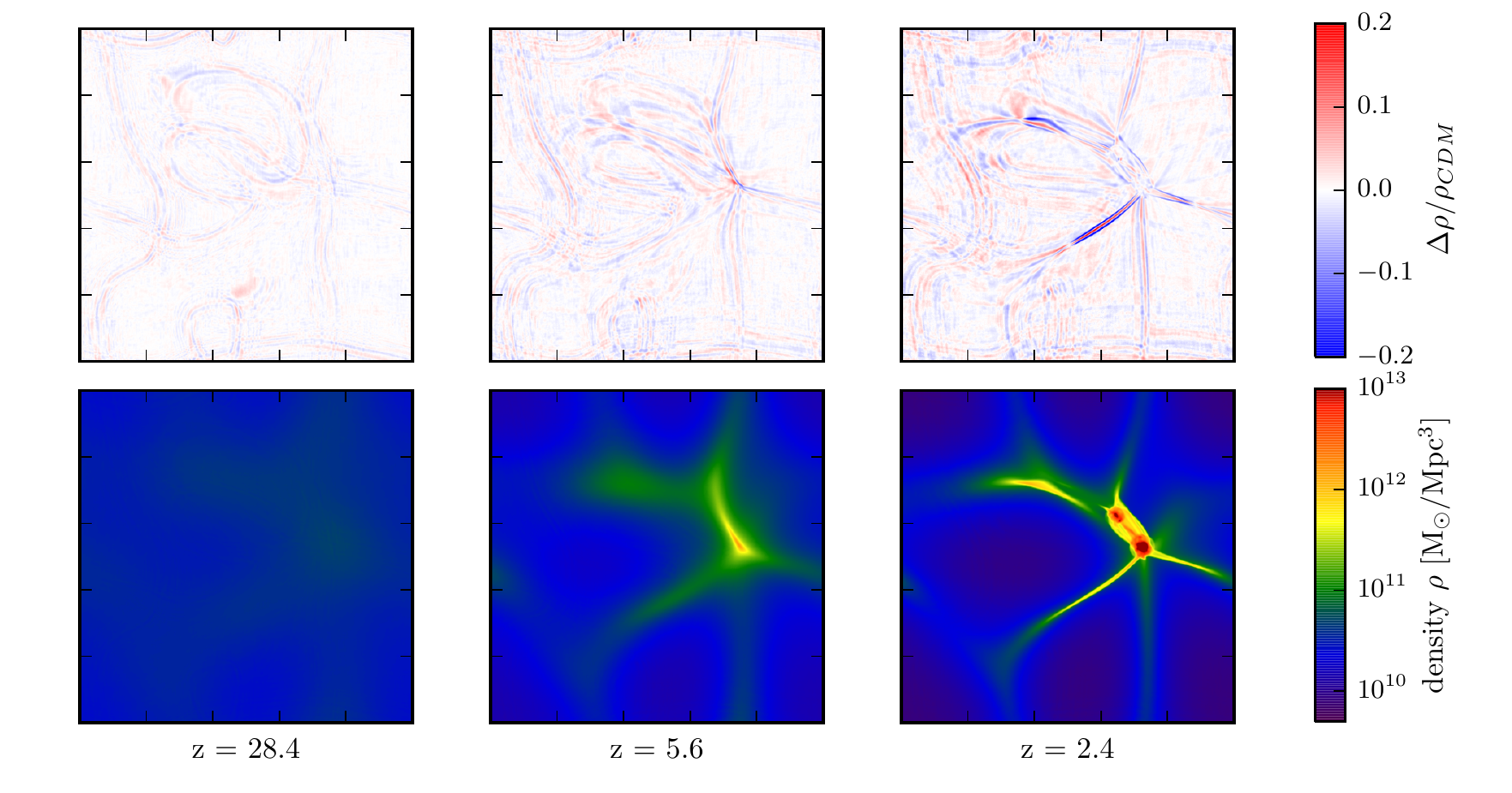}
 \caption{\textit{top:} Relative difference of simulations with quantum pressure compared to simulations without quantum pressure $(\rho_{FDM}-\rho_{CDM})/\rho_{CDM}$. \textit{bottom:} Density in the FDM simulation. The box size is $L = 2.5$ Mpc and the slice was chosen in a way that it intersects a massive halo.}
 \label{fig:plot7}
\end{figure*}

  \begin{figure}
\centering
 \includegraphics{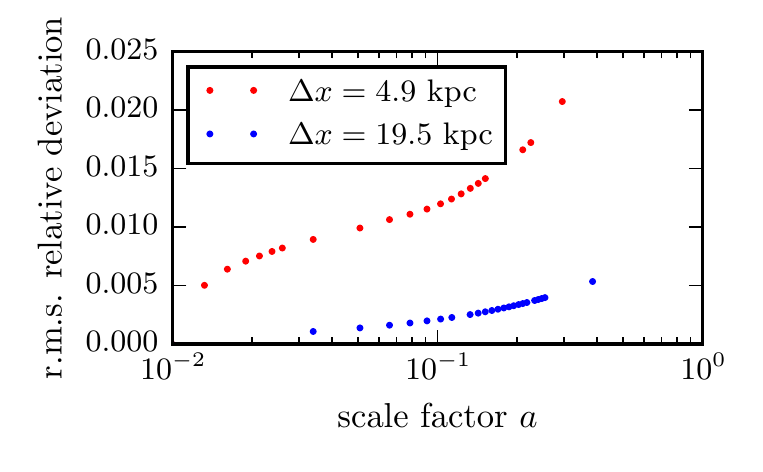}
 \caption{
 Root-mean-square relative difference of the density $(\rho_{FDM}-\rho_{CDM})/\rho_{CDM}$ per numerical cell for the two box sizes, i.e., cell widths of the numerical grid $\Delta x$.
 }
 \label{fig:plot8}
 \end{figure}

The relative difference of the FDM and CDM power spectra for both simulation boxes can be seen in \autoref{fig:plot6}. 
Our central result is that these differences grow up to $\approx 13 \%$ on scales close to the quantum Jeans length, while they become nearly undetectable on larger scales. Somewhat contrary to expectations, we observe more power on small scales in the FDM simulations compared to the pressureless N-body simulations. 
In both simulation boxes, the difference in the power spectrum drops again at the smallest resolved scales caused by the numerical particle smoothing.
Since we use an equal number of cells for both box sizes, the larger box corresponds to a lower physical resolution and the drop in the power spectrum due to particle smoothing occurs at larger scales. Both simulations are fully consistent apart from the grid cutoff. Based on these results alone, we cannot exclude an even larger difference in power on scales which are still unresolved in the smaller simulation box.
The results for the power spectrum are consistent with the relative density contrast whose amplitude is of the order of several $10\%$, as seen in \autoref{fig:plot7}. Interestingly, the largest differences are located around the filaments where an interference-like pattern can be identified. 

Finally, \autoref{fig:plot8} shows the root-mean-square relative deviations per cell. The deviations increase with decreasing redshift suggesting that even higher deviations can be found at $z=0$. Because differences exist primarily on smaller scales, the simulation with higher spatial resolution shows larger deviations.

 \section{Conclusions}
 \label{sec:conclusion}

We have presented a new particle-mesh scheme that allows the investigation of the effects of ultralight axion, or ``fuzzy'', dark matter (FDM) on cosmological structures at scales close to the quantum Jeans length. This was achieved by adding an additional force term, given by the quantum pressure in the Madelung form of the Schrödinger equation, to a standard N-body code in an energy conserving way. It was tested against known numerical results for the structure and oscillations of single solitonic cores, for the shapes of final radial density profiles in core mergers, and for the linear growth of perturbations in cosmological simulations. It is intended as an approximate method to find coarse-grained solutions to the governing Schrödinger-Poisson (SP) equations, capturing the resolved modifications to the density profile while leaving the small-scale interference pattern of the true SP solutions unresolved.

Our main goal was to detect the dynamical effects of quantum pressure during structure formation at lower redshifts, i.e., those that are not already captured by the linear transfer function used to initialize the simulations. Comparing our FDM simulations with the new scheme to standard N-body simulations with identical initial conditions, we indeed find significant differences in the resulting density fields. Interestingly, the largest relative deviations in the density field are located around the filaments, i.e. the first nonlinear structures to form on a given scale in cosmological structure formation (but note that our simulations fail to resolve solitonic halo cores where even larger differences might occur). The maximum difference in the resulting power spectra is of the order of $10\%$ at $z=2.38$ on scales around the quantum Jeans length. For observational probes of structure formation that reach a comparable level of precision, numerical predictions will have to include the dynamical effects of quantum pressure. This is our central result.

In future work, we will study the dynamical effects of FDM on the halo mass function and at lower redshifts, requiring larger simulation boxes while still resolving the Jeans length. It will also be interesting to search for potential observable consequences of the predicted density patterns around filaments.

\acknowledgements
We thank J.-F. Engels for help with the Nyx code and C. Behrens, X. Du, D.J.E. Marsh, and B. Schwabe for many useful discussions. The simulations were performed with resources provided by the North-German Supercomputing Alliance (HLRN). We acknowledge the yt toolkit \cite{Turk2011} that was used for our analysis of numerical data.


\begin{thebibliography}{43}%
\makeatletter
\providecommand \@ifxundefined [1]{%
 \@ifx{#1\undefined}
}%
\providecommand \@ifnum [1]{%
 \ifnum #1\expandafter \@firstoftwo
 \else \expandafter \@secondoftwo
 \fi
}%
\providecommand \@ifx [1]{%
 \ifx #1\expandafter \@firstoftwo
 \else \expandafter \@secondoftwo
 \fi
}%
\providecommand \natexlab [1]{#1}%
\providecommand \enquote  [1]{``#1''}%
\providecommand \bibnamefont  [1]{#1}%
\providecommand \bibfnamefont [1]{#1}%
\providecommand \citenamefont [1]{#1}%
\providecommand \href@noop [0]{\@secondoftwo}%
\providecommand \href [0]{\begingroup \@sanitize@url \@href}%
\providecommand \@href[1]{\@@startlink{#1}\@@href}%
\providecommand \@@href[1]{\endgroup#1\@@endlink}%
\providecommand \@sanitize@url [0]{\catcode `\\12\catcode `\$12\catcode
  `\&12\catcode `\#12\catcode `\^12\catcode `\_12\catcode `\%12\relax}%
\providecommand \@@startlink[1]{}%
\providecommand \@@endlink[0]{}%
\providecommand \url  [0]{\begingroup\@sanitize@url \@url }%
\providecommand \@url [1]{\endgroup\@href {#1}{\urlprefix }}%
\providecommand \urlprefix  [0]{URL }%
\providecommand \Eprint [0]{\href }%
\providecommand \doibase [0]{http://dx.doi.org/}%
\providecommand \selectlanguage [0]{\@gobble}%
\providecommand \bibinfo  [0]{\@secondoftwo}%
\providecommand \bibfield  [0]{\@secondoftwo}%
\providecommand \translation [1]{[#1]}%
\providecommand \BibitemOpen [0]{}%
\providecommand \bibitemStop [0]{}%
\providecommand \bibitemNoStop [0]{.\EOS\space}%
\providecommand \EOS [0]{\spacefactor3000\relax}%
\providecommand \BibitemShut  [1]{\csname bibitem#1\endcsname}%
\let\auto@bib@innerbib\@empty
\bibitem [{\citenamefont {Arvanitaki}\ \emph {et~al.}(2010)\citenamefont
  {Arvanitaki}, \citenamefont {Dimopoulos}, \citenamefont {Dubovsky},
  \citenamefont {Kaloper},\ and\ \citenamefont
  {March-Russell}}]{Arvanitaki2010}%
  \BibitemOpen
  \bibfield  {author} {\bibinfo {author} {\bibfnamefont {A.}~\bibnamefont
  {Arvanitaki}}, \bibinfo {author} {\bibfnamefont {S.}~\bibnamefont
  {Dimopoulos}}, \bibinfo {author} {\bibfnamefont {S.}~\bibnamefont
  {Dubovsky}}, \bibinfo {author} {\bibfnamefont {N.}~\bibnamefont {Kaloper}}, \
  and\ \bibinfo {author} {\bibfnamefont {J.}~\bibnamefont {March-Russell}},\
  }\href {\doibase 10.1103/PhysRevD.81.123530} {\bibfield  {journal} {\bibinfo
  {journal} {Phys. Rev. D}\ }\textbf {\bibinfo {volume} {81}},\ \bibinfo
  {pages} {123530} (\bibinfo {year} {2010})}\BibitemShut {NoStop}%
\bibitem [{\citenamefont {Hu}\ \emph {et~al.}(2000)\citenamefont {Hu},
  \citenamefont {Barkana},\ and\ \citenamefont {Gruzinov}}]{Hu2000}%
  \BibitemOpen
  \bibfield  {author} {\bibinfo {author} {\bibfnamefont {W.}~\bibnamefont
  {Hu}}, \bibinfo {author} {\bibfnamefont {R.}~\bibnamefont {Barkana}}, \ and\
  \bibinfo {author} {\bibfnamefont {A.}~\bibnamefont {Gruzinov}},\ }\href
  {\doibase 10.1103/physrevlett.85.1158} {\bibfield  {journal} {\bibinfo
  {journal} {Physical Review Letters}\ }\textbf {\bibinfo {volume} {85}},\
  \bibinfo {pages} {1158} (\bibinfo {year} {2000})}\BibitemShut {NoStop}%
\bibitem [{\citenamefont {{Su{\'a}rez}}\ and\ \citenamefont
  {{Matos}}(2011)}]{Suarez2011}%
  \BibitemOpen
  \bibfield  {author} {\bibinfo {author} {\bibfnamefont {A.}~\bibnamefont
  {{Su{\'a}rez}}}\ and\ \bibinfo {author} {\bibfnamefont {T.}~\bibnamefont
  {{Matos}}},\ }\href {\doibase 10.1111/j.1365-2966.2011.19012.x} {\bibfield
  {journal} {\bibinfo  {journal} {\mnras}\ }\textbf {\bibinfo {volume} {416}},\
  \bibinfo {pages} {87} (\bibinfo {year} {2011})},\ \Eprint
  {http://arxiv.org/abs/1101.4039} {arXiv:1101.4039 [gr-qc]} \BibitemShut
  {NoStop}%
\bibitem [{\citenamefont {{Robles}}\ and\ \citenamefont
  {{Matos}}(2012)}]{Robles2012}%
  \BibitemOpen
  \bibfield  {author} {\bibinfo {author} {\bibfnamefont {V.~H.}\ \bibnamefont
  {{Robles}}}\ and\ \bibinfo {author} {\bibfnamefont {T.}~\bibnamefont
  {{Matos}}},\ }\href {\doibase 10.1111/j.1365-2966.2012.20603.x} {\bibfield
  {journal} {\bibinfo  {journal} {\mnras}\ }\textbf {\bibinfo {volume} {422}},\
  \bibinfo {pages} {282} (\bibinfo {year} {2012})},\ \Eprint
  {http://arxiv.org/abs/1201.3032} {arXiv:1201.3032 [astro-ph.CO]} \BibitemShut
  {NoStop}%
\bibitem [{\citenamefont {{Rindler-Daller}}\ and\ \citenamefont
  {{Shapiro}}(2014)}]{Rindler-Daller2014}%
  \BibitemOpen
  \bibfield  {author} {\bibinfo {author} {\bibfnamefont {T.}~\bibnamefont
  {{Rindler-Daller}}}\ and\ \bibinfo {author} {\bibfnamefont {P.~R.}\
  \bibnamefont {{Shapiro}}},\ }\href {\doibase 10.1142/S021773231430002X}
  {\bibfield  {journal} {\bibinfo  {journal} {Modern Physics Letters A}\
  }\textbf {\bibinfo {volume} {29}},\ \bibinfo {eid} {1430002} (\bibinfo {year}
  {2014})},\ \Eprint {http://arxiv.org/abs/1312.1734} {arXiv:1312.1734}
  \BibitemShut {NoStop}%
\bibitem [{\citenamefont {{Lora}}\ and\ \citenamefont
  {{Maga{\~n}a}}(2014)}]{Lora2014}%
  \BibitemOpen
  \bibfield  {author} {\bibinfo {author} {\bibfnamefont {V.}~\bibnamefont
  {{Lora}}}\ and\ \bibinfo {author} {\bibfnamefont {J.}~\bibnamefont
  {{Maga{\~n}a}}},\ }\href {\doibase 10.1088/1475-7516/2014/09/011} {\bibfield
  {journal} {\bibinfo  {journal} {\jcap}\ }\textbf {\bibinfo {volume} {9}},\
  \bibinfo {eid} {011} (\bibinfo {year} {2014})},\ \Eprint
  {http://arxiv.org/abs/1406.6875} {arXiv:1406.6875} \BibitemShut {NoStop}%
\bibitem [{\citenamefont {{Martinez-Medina}}\ \emph {et~al.}(2015)\citenamefont
  {{Martinez-Medina}}, \citenamefont {{Robles}},\ and\ \citenamefont
  {{Matos}}}]{Martinez-Medina2015}%
  \BibitemOpen
  \bibfield  {author} {\bibinfo {author} {\bibfnamefont {L.~A.}\ \bibnamefont
  {{Martinez-Medina}}}, \bibinfo {author} {\bibfnamefont {V.~H.}\ \bibnamefont
  {{Robles}}}, \ and\ \bibinfo {author} {\bibfnamefont {T.}~\bibnamefont
  {{Matos}}},\ }\href {\doibase 10.1103/PhysRevD.91.023519} {\bibfield
  {journal} {\bibinfo  {journal} {\prd}\ }\textbf {\bibinfo {volume} {91}},\
  \bibinfo {eid} {023519} (\bibinfo {year} {2015})},\ \Eprint
  {http://arxiv.org/abs/1410.4163} {arXiv:1410.4163} \BibitemShut {NoStop}%
\bibitem [{\citenamefont {{Chavanis}}(2011)}]{Chavanis2011}%
  \BibitemOpen
  \bibfield  {author} {\bibinfo {author} {\bibfnamefont {P.-H.}\ \bibnamefont
  {{Chavanis}}},\ }\href {\doibase 10.1103/PhysRevD.84.043531} {\bibfield
  {journal} {\bibinfo  {journal} {\prd}\ }\textbf {\bibinfo {volume} {84}},\
  \bibinfo {eid} {043531} (\bibinfo {year} {2011})},\ \Eprint
  {http://arxiv.org/abs/1103.2050} {arXiv:1103.2050} \BibitemShut {NoStop}%
\bibitem [{\citenamefont {{Guzm{\'a}n}}\ and\ \citenamefont
  {{Lora-Clavijo}}(2015)}]{Guzman2015}%
  \BibitemOpen
  \bibfield  {author} {\bibinfo {author} {\bibfnamefont {F.~S.}\ \bibnamefont
  {{Guzm{\'a}n}}}\ and\ \bibinfo {author} {\bibfnamefont {F.~D.}\ \bibnamefont
  {{Lora-Clavijo}}},\ }\href {\doibase 10.1007/s10714-015-1865-9} {\bibfield
  {journal} {\bibinfo  {journal} {General Relativity and Gravitation}\ }\textbf
  {\bibinfo {volume} {47}},\ \bibinfo {eid} {21} (\bibinfo {year} {2015})},\
  \Eprint {http://arxiv.org/abs/1501.06553} {arXiv:1501.06553} \BibitemShut
  {NoStop}%
\bibitem [{\citenamefont {{Lee}}(2016)}]{Lee2016}%
  \BibitemOpen
  \bibfield  {author} {\bibinfo {author} {\bibfnamefont {J.-W.}\ \bibnamefont
  {{Lee}}},\ }\href {\doibase 10.1016/j.physletb.2016.03.016} {\bibfield
  {journal} {\bibinfo  {journal} {Physics Letters B}\ }\textbf {\bibinfo
  {volume} {756}},\ \bibinfo {pages} {166} (\bibinfo {year} {2016})},\ \Eprint
  {http://arxiv.org/abs/1511.06611} {arXiv:1511.06611} \BibitemShut {NoStop}%
\bibitem [{\citenamefont {{Chen}}\ \emph {et~al.}(2016)\citenamefont {{Chen}},
  \citenamefont {{Schive}},\ and\ \citenamefont {{Chiueh}}}]{Chen2016}%
  \BibitemOpen
  \bibfield  {author} {\bibinfo {author} {\bibfnamefont {S.-R.}\ \bibnamefont
  {{Chen}}}, \bibinfo {author} {\bibfnamefont {H.-Y.}\ \bibnamefont
  {{Schive}}}, \ and\ \bibinfo {author} {\bibfnamefont {T.}~\bibnamefont
  {{Chiueh}}},\ }\href@noop {} {\bibfield  {journal} {\bibinfo  {journal}
  {ArXiv e-prints}\ } (\bibinfo {year} {2016})},\ \Eprint
  {http://arxiv.org/abs/1606.09030} {arXiv:1606.09030} \BibitemShut {NoStop}%
\bibitem [{\citenamefont {{Marsh}}(2016)}]{Marsh2015a}%
  \BibitemOpen
  \bibfield  {author} {\bibinfo {author} {\bibfnamefont {D.~J.~E.}\
  \bibnamefont {{Marsh}}},\ }\href {\doibase 10.1016/j.physrep.2016.06.005}
  {\bibfield  {journal} {\bibinfo  {journal} {Physics Reports}\ }\textbf
  {\bibinfo {volume} {643}},\ \bibinfo {pages} {1} (\bibinfo {year} {2016})},\
  \Eprint {http://arxiv.org/abs/1510.07633} {arXiv:1510.07633} \BibitemShut
  {NoStop}%
\bibitem [{\citenamefont {Peebles}(2000)}]{Peebles2000}%
  \BibitemOpen
  \bibfield  {author} {\bibinfo {author} {\bibfnamefont {P.~J.~E.}\
  \bibnamefont {Peebles}},\ }\href
  {http://stacks.iop.org/1538-4357/534/i=2/a=L127} {\bibfield  {journal}
  {\bibinfo  {journal} {The Astrophysical Journal Letters}\ }\textbf {\bibinfo
  {volume} {534}},\ \bibinfo {pages} {L127} (\bibinfo {year}
  {2000})}\BibitemShut {NoStop}%
\bibitem [{\citenamefont {{Marsh}}\ and\ \citenamefont
  {{Silk}}(2014)}]{Marsh2014}%
  \BibitemOpen
  \bibfield  {author} {\bibinfo {author} {\bibfnamefont {D.~J.~E.}\
  \bibnamefont {{Marsh}}}\ and\ \bibinfo {author} {\bibfnamefont
  {J.}~\bibnamefont {{Silk}}},\ }\href {\doibase 10.1093/mnras/stt2079}
  {\bibfield  {journal} {\bibinfo  {journal} {\mnras}\ }\textbf {\bibinfo
  {volume} {437}},\ \bibinfo {pages} {2652} (\bibinfo {year} {2014})},\ \Eprint
  {http://arxiv.org/abs/1307.1705} {arXiv:1307.1705 [astro-ph.CO]} \BibitemShut
  {NoStop}%
\bibitem [{\citenamefont {Moore}\ \emph {et~al.}(1999)\citenamefont {Moore},
  \citenamefont {Ghigna}, \citenamefont {Governato}, \citenamefont {Lake},
  \citenamefont {Quinn}, \citenamefont {Stadel},\ and\ \citenamefont
  {Tozzi}}]{Moore1999}%
  \BibitemOpen
  \bibfield  {author} {\bibinfo {author} {\bibfnamefont {B.}~\bibnamefont
  {Moore}}, \bibinfo {author} {\bibfnamefont {S.}~\bibnamefont {Ghigna}},
  \bibinfo {author} {\bibfnamefont {F.}~\bibnamefont {Governato}}, \bibinfo
  {author} {\bibfnamefont {G.}~\bibnamefont {Lake}}, \bibinfo {author}
  {\bibfnamefont {T.}~\bibnamefont {Quinn}}, \bibinfo {author} {\bibfnamefont
  {J.}~\bibnamefont {Stadel}}, \ and\ \bibinfo {author} {\bibfnamefont
  {P.}~\bibnamefont {Tozzi}},\ }\href {\doibase 10.1086/312287} {\bibfield
  {journal} {\bibinfo  {journal} {\apj}\ }\textbf {\bibinfo {volume} {524}},\
  \bibinfo {pages} {L19} (\bibinfo {year} {1999})}\BibitemShut {NoStop}%
\bibitem [{\citenamefont {Klypin}\ \emph {et~al.}(1999)\citenamefont {Klypin},
  \citenamefont {Kravtsov}, \citenamefont {Valenzuela},\ and\ \citenamefont
  {Prada}}]{Klypin1999}%
  \BibitemOpen
  \bibfield  {author} {\bibinfo {author} {\bibfnamefont {A.}~\bibnamefont
  {Klypin}}, \bibinfo {author} {\bibfnamefont {A.~V.}\ \bibnamefont
  {Kravtsov}}, \bibinfo {author} {\bibfnamefont {O.}~\bibnamefont
  {Valenzuela}}, \ and\ \bibinfo {author} {\bibfnamefont {F.}~\bibnamefont
  {Prada}},\ }\href {http://stacks.iop.org/0004-637X/522/i=1/a=82} {\bibfield
  {journal} {\bibinfo  {journal} {\apj}\ }\textbf {\bibinfo {volume} {522}},\
  \bibinfo {pages} {82} (\bibinfo {year} {1999})}\BibitemShut {NoStop}%
\bibitem [{\citenamefont {{Gilmore}}\ \emph {et~al.}(2007)\citenamefont
  {{Gilmore}}, \citenamefont {{Wilkinson}}, \citenamefont {{Wyse}},
  \citenamefont {{Kleyna}}, \citenamefont {{Koch}}, \citenamefont {{Evans}},\
  and\ \citenamefont {{Grebel}}}]{Gilmore2007}%
  \BibitemOpen
  \bibfield  {author} {\bibinfo {author} {\bibfnamefont {G.}~\bibnamefont
  {{Gilmore}}}, \bibinfo {author} {\bibfnamefont {M.~I.}\ \bibnamefont
  {{Wilkinson}}}, \bibinfo {author} {\bibfnamefont {R.~F.~G.}\ \bibnamefont
  {{Wyse}}}, \bibinfo {author} {\bibfnamefont {J.~T.}\ \bibnamefont
  {{Kleyna}}}, \bibinfo {author} {\bibfnamefont {A.}~\bibnamefont {{Koch}}},
  \bibinfo {author} {\bibfnamefont {N.~W.}\ \bibnamefont {{Evans}}}, \ and\
  \bibinfo {author} {\bibfnamefont {E.~K.}\ \bibnamefont {{Grebel}}},\ }\href
  {\doibase 10.1086/518025} {\bibfield  {journal} {\bibinfo  {journal} {\apj}\
  }\textbf {\bibinfo {volume} {663}},\ \bibinfo {pages} {948} (\bibinfo {year}
  {2007})},\ \Eprint {http://arxiv.org/abs/astro-ph/0703308} {astro-ph/0703308}
  \BibitemShut {NoStop}%
\bibitem [{\citenamefont {{Arvanitaki}}\ \emph {et~al.}(2015)\citenamefont
  {{Arvanitaki}}, \citenamefont {{Baryakhtar}},\ and\ \citenamefont
  {{Huang}}}]{Arvanitaki2015}%
  \BibitemOpen
  \bibfield  {author} {\bibinfo {author} {\bibfnamefont {A.}~\bibnamefont
  {{Arvanitaki}}}, \bibinfo {author} {\bibfnamefont {M.}~\bibnamefont
  {{Baryakhtar}}}, \ and\ \bibinfo {author} {\bibfnamefont {X.}~\bibnamefont
  {{Huang}}},\ }\href {\doibase 10.1103/PhysRevD.91.084011} {\bibfield
  {journal} {\bibinfo  {journal} {\prd}\ }\textbf {\bibinfo {volume} {91}},\
  \bibinfo {eid} {084011} (\bibinfo {year} {2015})},\ \Eprint
  {http://arxiv.org/abs/1411.2263} {arXiv:1411.2263 [hep-ph]} \BibitemShut
  {NoStop}%
\bibitem [{\citenamefont {{Bozek}}\ \emph {et~al.}(2015)\citenamefont
  {{Bozek}}, \citenamefont {{Marsh}}, \citenamefont {{Silk}},\ and\
  \citenamefont {{Wyse}}}]{Bozek2015}%
  \BibitemOpen
  \bibfield  {author} {\bibinfo {author} {\bibfnamefont {B.}~\bibnamefont
  {{Bozek}}}, \bibinfo {author} {\bibfnamefont {D.~J.~E.}\ \bibnamefont
  {{Marsh}}}, \bibinfo {author} {\bibfnamefont {J.}~\bibnamefont {{Silk}}}, \
  and\ \bibinfo {author} {\bibfnamefont {R.~F.~G.}\ \bibnamefont {{Wyse}}},\
  }\href {\doibase 10.1093/mnras/stv624} {\bibfield  {journal} {\bibinfo
  {journal} {\mnras}\ }\textbf {\bibinfo {volume} {450}},\ \bibinfo {pages}
  {209} (\bibinfo {year} {2015})},\ \Eprint {http://arxiv.org/abs/1409.3544}
  {arXiv:1409.3544} \BibitemShut {NoStop}%
\bibitem [{\citenamefont {Schive}\ \emph {et~al.}(2016)\citenamefont {Schive},
  \citenamefont {Chiueh}, \citenamefont {Broadhurst},\ and\ \citenamefont
  {Huang}}]{Schive2016}%
  \BibitemOpen
  \bibfield  {author} {\bibinfo {author} {\bibfnamefont {H.-Y.}\ \bibnamefont
  {Schive}}, \bibinfo {author} {\bibfnamefont {T.}~\bibnamefont {Chiueh}},
  \bibinfo {author} {\bibfnamefont {T.}~\bibnamefont {Broadhurst}}, \ and\
  \bibinfo {author} {\bibfnamefont {K.-W.}\ \bibnamefont {Huang}},\ }\href
  {\doibase 10.3847/0004-637x/818/1/89} {\bibfield  {journal} {\bibinfo
  {journal} {\apj}\ }\textbf {\bibinfo {volume} {818}},\ \bibinfo {pages} {89}
  (\bibinfo {year} {2016})}\BibitemShut {NoStop}%
\bibitem [{\citenamefont {Sarkar}\ \emph {et~al.}(2016)\citenamefont {Sarkar},
  \citenamefont {Mondal}, \citenamefont {Das}, \citenamefont {Sethi},
  \citenamefont {Bharadwaj},\ and\ \citenamefont {Marsh}}]{Sarkar2016}%
  \BibitemOpen
  \bibfield  {author} {\bibinfo {author} {\bibfnamefont {A.}~\bibnamefont
  {Sarkar}}, \bibinfo {author} {\bibfnamefont {R.}~\bibnamefont {Mondal}},
  \bibinfo {author} {\bibfnamefont {S.}~\bibnamefont {Das}}, \bibinfo {author}
  {\bibfnamefont {S.}~\bibnamefont {Sethi}}, \bibinfo {author} {\bibfnamefont
  {S.}~\bibnamefont {Bharadwaj}}, \ and\ \bibinfo {author} {\bibfnamefont
  {D.~J.}\ \bibnamefont {Marsh}},\ }\href
  {http://stacks.iop.org/1475-7516/2016/i=04/a=012} {\bibfield  {journal}
  {\bibinfo  {journal} {\jcap}\ }\textbf {\bibinfo {volume} {04}},\ \bibinfo
  {pages} {012} (\bibinfo {year} {2016})}\BibitemShut {NoStop}%
\bibitem [{\citenamefont {Schive}\ \emph
  {et~al.}(2014{\natexlab{a}})\citenamefont {Schive}, \citenamefont {Liao},
  \citenamefont {Woo}, \citenamefont {Wong}, \citenamefont {Chiueh},
  \citenamefont {Broadhurst},\ and\ \citenamefont {Hwang}}]{Schive2014a}%
  \BibitemOpen
  \bibfield  {author} {\bibinfo {author} {\bibfnamefont {H.-Y.}\ \bibnamefont
  {Schive}}, \bibinfo {author} {\bibfnamefont {M.-H.}\ \bibnamefont {Liao}},
  \bibinfo {author} {\bibfnamefont {T.-P.}\ \bibnamefont {Woo}}, \bibinfo
  {author} {\bibfnamefont {S.-K.}\ \bibnamefont {Wong}}, \bibinfo {author}
  {\bibfnamefont {T.}~\bibnamefont {Chiueh}}, \bibinfo {author} {\bibfnamefont
  {T.}~\bibnamefont {Broadhurst}}, \ and\ \bibinfo {author} {\bibfnamefont
  {W.-Y.~P.}\ \bibnamefont {Hwang}},\ }\href {\doibase
  10.1103/PhysRevLett.113.261302} {\bibfield  {journal} {\bibinfo  {journal}
  {Phys. Rev. Lett.}\ }\textbf {\bibinfo {volume} {113}},\ \bibinfo {pages}
  {261302} (\bibinfo {year} {2014}{\natexlab{a}})}\BibitemShut {NoStop}%
\bibitem [{\citenamefont {{Marsh}}\ and\ \citenamefont
  {{Pop}}(2015)}]{Marsh2015c}%
  \BibitemOpen
  \bibfield  {author} {\bibinfo {author} {\bibfnamefont {D.~J.~E.}\
  \bibnamefont {{Marsh}}}\ and\ \bibinfo {author} {\bibfnamefont {A.-R.}\
  \bibnamefont {{Pop}}},\ }\href {\doibase 10.1093/mnras/stv1050} {\bibfield
  {journal} {\bibinfo  {journal} {\mnras}\ }\textbf {\bibinfo {volume} {451}},\
  \bibinfo {pages} {2479} (\bibinfo {year} {2015})},\ \Eprint
  {http://arxiv.org/abs/1502.03456} {arXiv:1502.03456} \BibitemShut {NoStop}%
\bibitem [{\citenamefont {{Calabrese}}\ and\ \citenamefont
  {{Spergel}}(2016)}]{Calabrese2016}%
  \BibitemOpen
  \bibfield  {author} {\bibinfo {author} {\bibfnamefont {E.}~\bibnamefont
  {{Calabrese}}}\ and\ \bibinfo {author} {\bibfnamefont {D.~N.}\ \bibnamefont
  {{Spergel}}},\ }\href {\doibase 10.1093/mnras/stw1256} {\bibfield  {journal}
  {\bibinfo  {journal} {\mnras}\ }\textbf {\bibinfo {volume} {460}},\ \bibinfo
  {pages} {4397} (\bibinfo {year} {2016})},\ \Eprint
  {http://arxiv.org/abs/1603.07321} {arXiv:1603.07321} \BibitemShut {NoStop}%
\bibitem [{\citenamefont {{Macci{\`o}}}\ \emph {et~al.}(2012)\citenamefont
  {{Macci{\`o}}}, \citenamefont {{Paduroiu}}, \citenamefont {{Anderhalden}},
  \citenamefont {{Schneider}},\ and\ \citenamefont {{Moore}}}]{Maccio2012}%
  \BibitemOpen
  \bibfield  {author} {\bibinfo {author} {\bibfnamefont {A.~V.}\ \bibnamefont
  {{Macci{\`o}}}}, \bibinfo {author} {\bibfnamefont {S.}~\bibnamefont
  {{Paduroiu}}}, \bibinfo {author} {\bibfnamefont {D.}~\bibnamefont
  {{Anderhalden}}}, \bibinfo {author} {\bibfnamefont {A.}~\bibnamefont
  {{Schneider}}}, \ and\ \bibinfo {author} {\bibfnamefont {B.}~\bibnamefont
  {{Moore}}},\ }\href {\doibase 10.1111/j.1365-2966.2012.21284.x} {\bibfield
  {journal} {\bibinfo  {journal} {\mnras}\ }\textbf {\bibinfo {volume} {424}},\
  \bibinfo {pages} {1105} (\bibinfo {year} {2012})},\ \Eprint
  {http://arxiv.org/abs/1202.1282} {arXiv:1202.1282} \BibitemShut {NoStop}%
\bibitem [{\citenamefont {Ruffini}\ and\ \citenamefont
  {Bonazzola}(1969)}]{Ruffini}%
  \BibitemOpen
  \bibfield  {author} {\bibinfo {author} {\bibfnamefont {R.}~\bibnamefont
  {Ruffini}}\ and\ \bibinfo {author} {\bibfnamefont {S.}~\bibnamefont
  {Bonazzola}},\ }\href {\doibase 10.1103/PhysRev.187.1767} {\bibfield
  {journal} {\bibinfo  {journal} {Phys. Rev.}\ }\textbf {\bibinfo {volume}
  {187}},\ \bibinfo {pages} {1767} (\bibinfo {year} {1969})}\BibitemShut
  {NoStop}%
\bibitem [{\citenamefont {Guzm\'an}\ and\ \citenamefont
  {Ure{\~{n}}a-L{\'{o}}pez}(2004)}]{Guzman2004}%
  \BibitemOpen
  \bibfield  {author} {\bibinfo {author} {\bibfnamefont {F.~S.}\ \bibnamefont
  {Guzm\'an}}\ and\ \bibinfo {author} {\bibfnamefont {L.~A.}\ \bibnamefont
  {Ure{\~{n}}a-L{\'{o}}pez}},\ }\href {\doibase 10.1103/PhysRevD.69.124033}
  {\bibfield  {journal} {\bibinfo  {journal} {Phys. Rev. D}\ }\textbf {\bibinfo
  {volume} {69}},\ \bibinfo {pages} {124033} (\bibinfo {year}
  {2004})}\BibitemShut {NoStop}%
\bibitem [{\citenamefont {Bernal}\ and\ \citenamefont
  {Guzm\'an}(2006{\natexlab{a}})}]{Bernal2006a}%
  \BibitemOpen
  \bibfield  {author} {\bibinfo {author} {\bibfnamefont {A.}~\bibnamefont
  {Bernal}}\ and\ \bibinfo {author} {\bibfnamefont {F.~S.}\ \bibnamefont
  {Guzm\'an}},\ }\href {\doibase 10.1103/PhysRevD.74.063504} {\bibfield
  {journal} {\bibinfo  {journal} {Phys. Rev. D}\ }\textbf {\bibinfo {volume}
  {74}},\ \bibinfo {pages} {063504} (\bibinfo {year}
  {2006}{\natexlab{a}})}\BibitemShut {NoStop}%
\bibitem [{\citenamefont {Bernal}\ and\ \citenamefont
  {Guzm\'an}(2006{\natexlab{b}})}]{Bernal2006}%
  \BibitemOpen
  \bibfield  {author} {\bibinfo {author} {\bibfnamefont {A.}~\bibnamefont
  {Bernal}}\ and\ \bibinfo {author} {\bibfnamefont {F.~S.}\ \bibnamefont
  {Guzm\'an}},\ }\href {\doibase 10.1103/PhysRevD.74.103002} {\bibfield
  {journal} {\bibinfo  {journal} {Phys. Rev. D}\ }\textbf {\bibinfo {volume}
  {74}},\ \bibinfo {pages} {103002} (\bibinfo {year}
  {2006}{\natexlab{b}})}\BibitemShut {NoStop}%
\bibitem [{\citenamefont {Liebling}\ and\ \citenamefont
  {Palenzuela}(2012)}]{Liebling2012}%
  \BibitemOpen
  \bibfield  {author} {\bibinfo {author} {\bibfnamefont {S.~L.}\ \bibnamefont
  {Liebling}}\ and\ \bibinfo {author} {\bibfnamefont {C.}~\bibnamefont
  {Palenzuela}},\ }\href {\doibase 10.12942/lrr-2012-6} {\bibfield  {journal}
  {\bibinfo  {journal} {Living Reviews in Relativity}\ }\textbf {\bibinfo
  {volume} {15}},\ \bibinfo {pages} {1} (\bibinfo {year} {2012})},\ \Eprint
  {http://arxiv.org/abs/1202.5809} {arXiv:1202.5809} \BibitemShut {NoStop}%
\bibitem [{\citenamefont {{Gonz{\'a}lez}}\ and\ \citenamefont
  {{Guzm{\'a}n}}(2011)}]{Gonzales2011}%
  \BibitemOpen
  \bibfield  {author} {\bibinfo {author} {\bibfnamefont {J.~A.}\ \bibnamefont
  {{Gonz{\'a}lez}}}\ and\ \bibinfo {author} {\bibfnamefont {F.~S.}\
  \bibnamefont {{Guzm{\'a}n}}},\ }\href {\doibase 10.1103/PhysRevD.83.103513}
  {\bibfield  {journal} {\bibinfo  {journal} {\prd}\ }\textbf {\bibinfo
  {volume} {83}},\ \bibinfo {eid} {103513} (\bibinfo {year} {2011})},\ \Eprint
  {http://arxiv.org/abs/1105.2066} {arXiv:1105.2066 [astro-ph.CO]} \BibitemShut
  {NoStop}%
\bibitem [{\citenamefont {{Woo}}\ and\ \citenamefont
  {{Chiueh}}(2009)}]{Woo2009}%
  \BibitemOpen
  \bibfield  {author} {\bibinfo {author} {\bibfnamefont {T.-P.}\ \bibnamefont
  {{Woo}}}\ and\ \bibinfo {author} {\bibfnamefont {T.}~\bibnamefont
  {{Chiueh}}},\ }\href {\doibase 10.1088/0004-637X/697/1/850} {\bibfield
  {journal} {\bibinfo  {journal} {\apj}\ }\textbf {\bibinfo {volume} {697}},\
  \bibinfo {pages} {850} (\bibinfo {year} {2009})},\ \Eprint
  {http://arxiv.org/abs/0806.0232} {arXiv:0806.0232} \BibitemShut {NoStop}%
\bibitem [{\citenamefont {Schive}\ \emph
  {et~al.}(2014{\natexlab{b}})\citenamefont {Schive}, \citenamefont {Chiueh},\
  and\ \citenamefont {Broadhurst}}]{Schive2014}%
  \BibitemOpen
  \bibfield  {author} {\bibinfo {author} {\bibfnamefont {H.-Y.}\ \bibnamefont
  {Schive}}, \bibinfo {author} {\bibfnamefont {T.}~\bibnamefont {Chiueh}}, \
  and\ \bibinfo {author} {\bibfnamefont {T.}~\bibnamefont {Broadhurst}},\
  }\href@noop {} {\bibfield  {journal} {\bibinfo  {journal} {Nature Physics}\
  }\textbf {\bibinfo {volume} {10}},\ \bibinfo {pages} {496} (\bibinfo {year}
  {2014}{\natexlab{b}})},\ \Eprint {http://arxiv.org/abs/1406.6586} {1406.6586}
  \BibitemShut {NoStop}%
\bibitem [{\citenamefont {{Schwabe}}\ \emph {et~al.}(2016)\citenamefont
  {{Schwabe}}, \citenamefont {{Niemeyer}},\ and\ \citenamefont
  {{Engels}}}]{Schwabe2016}%
  \BibitemOpen
  \bibfield  {author} {\bibinfo {author} {\bibfnamefont {B.}~\bibnamefont
  {{Schwabe}}}, \bibinfo {author} {\bibfnamefont {J.~C.}\ \bibnamefont
  {{Niemeyer}}}, \ and\ \bibinfo {author} {\bibfnamefont {J.~F.}\ \bibnamefont
  {{Engels}}},\ }\href {\doibase 10.1103/PhysRevD.94.043513} {\bibfield
  {journal} {\bibinfo  {journal} {\prd}\ }\textbf {\bibinfo {volume} {94}},\
  \bibinfo {eid} {043513} (\bibinfo {year} {2016})},\ \Eprint
  {http://arxiv.org/abs/1606.05151} {arXiv:1606.05151} \BibitemShut {NoStop}%
\bibitem [{\citenamefont {Mocz}\ and\ \citenamefont {Succi}(2015)}]{Mocz2015}%
  \BibitemOpen
  \bibfield  {author} {\bibinfo {author} {\bibfnamefont {P.}~\bibnamefont
  {Mocz}}\ and\ \bibinfo {author} {\bibfnamefont {S.}~\bibnamefont {Succi}},\
  }\href {\doibase 10.1103/PhysRevE.91.053304} {\bibfield  {journal} {\bibinfo
  {journal} {Physical Review E}\ }\textbf {\bibinfo {volume} {91}},\ \bibinfo
  {pages} {053304} (\bibinfo {year} {2015})}\BibitemShut {NoStop}%
\bibitem [{\citenamefont {Marsh}(2015)}]{Marsh2015b}%
  \BibitemOpen
  \bibfield  {author} {\bibinfo {author} {\bibfnamefont {D.~J.~E.}\
  \bibnamefont {Marsh}},\ }\href {\doibase 10.1103/PhysRevD.91.123520}
  {\bibfield  {journal} {\bibinfo  {journal} {Phys. Rev. D}\ }\textbf {\bibinfo
  {volume} {91}},\ \bibinfo {pages} {123520} (\bibinfo {year}
  {2015})}\BibitemShut {NoStop}%
\bibitem [{\citenamefont {Madelung}(1927)}]{Madelung1927}%
  \BibitemOpen
  \bibfield  {author} {\bibinfo {author} {\bibfnamefont {E.}~\bibnamefont
  {Madelung}},\ }\href {\doibase 10.1007/BF01400372} {\bibfield  {journal}
  {\bibinfo  {journal} {Zeitschrift für Physik}\ }\textbf {\bibinfo {volume}
  {40}},\ \bibinfo {pages} {322} (\bibinfo {year} {1927})}\BibitemShut
  {NoStop}%
\bibitem [{\citenamefont {Almgren}\ \emph {et~al.}(2013)\citenamefont
  {Almgren}, \citenamefont {Bell}, \citenamefont {Lijewski}, \citenamefont
  {Lukic},\ and\ \citenamefont {Van~Andel}}]{Almgren2013}%
  \BibitemOpen
  \bibfield  {author} {\bibinfo {author} {\bibfnamefont {A.~S.}\ \bibnamefont
  {Almgren}}, \bibinfo {author} {\bibfnamefont {J.~B.}\ \bibnamefont {Bell}},
  \bibinfo {author} {\bibfnamefont {M.~J.}\ \bibnamefont {Lijewski}}, \bibinfo
  {author} {\bibfnamefont {Z.}~\bibnamefont {Lukic}}, \ and\ \bibinfo {author}
  {\bibfnamefont {E.}~\bibnamefont {Van~Andel}},\ }\href {\doibase
  10.1088/0004-637x/765/1/39} {\bibfield  {journal} {\bibinfo  {journal}
  {\apj}\ }\textbf {\bibinfo {volume} {765}},\ \bibinfo {pages} {39} (\bibinfo
  {year} {2013})}\BibitemShut {NoStop}%
\bibitem [{\citenamefont {Monaghan}(2005)}]{Monaghan2005}%
  \BibitemOpen
  \bibfield  {author} {\bibinfo {author} {\bibfnamefont {J.~J.}\ \bibnamefont
  {Monaghan}},\ }\href {\doibase 10.1088/0034-4885/68/8/r01} {\bibfield
  {journal} {\bibinfo  {journal} {Reports on Progress in Physics}\ }\textbf
  {\bibinfo {volume} {68}},\ \bibinfo {pages} {1703} (\bibinfo {year}
  {2005})}\BibitemShut {NoStop}%
\bibitem [{\citenamefont {Monaghan}(1992)}]{Monaghan1992}%
  \BibitemOpen
  \bibfield  {author} {\bibinfo {author} {\bibfnamefont {J.~J.}\ \bibnamefont
  {Monaghan}},\ }\href {\doibase 10.1146/annurev.aa.30.090192.002551}
  {\bibfield  {journal} {\bibinfo  {journal} {Annual Review of Astronomy and
  Astrophysics}\ }\textbf {\bibinfo {volume} {30}},\ \bibinfo {pages} {543}
  (\bibinfo {year} {1992})}\BibitemShut {NoStop}%
\bibitem [{\citenamefont {Hahn}\ and\ \citenamefont {Abel}(2011)}]{Hahn2011}%
  \BibitemOpen
  \bibfield  {author} {\bibinfo {author} {\bibfnamefont {O.}~\bibnamefont
  {Hahn}}\ and\ \bibinfo {author} {\bibfnamefont {T.}~\bibnamefont {Abel}},\
  }\href {\doibase 10.1111/j.1365-2966.2011.18820.x} {\bibfield  {journal}
  {\bibinfo  {journal} {\mnras}\ }\textbf {\bibinfo {volume} {415}},\ \bibinfo
  {pages} {2101} (\bibinfo {year} {2011})}\BibitemShut {NoStop}%
\bibitem [{\citenamefont {Eisenstein}\ and\ \citenamefont
  {Hu}(1998)}]{Eisenstein1998}%
  \BibitemOpen
  \bibfield  {author} {\bibinfo {author} {\bibfnamefont {D.~J.}\ \bibnamefont
  {Eisenstein}}\ and\ \bibinfo {author} {\bibfnamefont {W.}~\bibnamefont
  {Hu}},\ }\href {http://stacks.iop.org/0004-637X/496/i=2/a=605} {\bibfield
  {journal} {\bibinfo  {journal} {\apj}\ }\textbf {\bibinfo {volume} {496}},\
  \bibinfo {pages} {605} (\bibinfo {year} {1998})}\BibitemShut {NoStop}%
\bibitem [{\citenamefont {{Turk}}\ \emph {et~al.}(2011)\citenamefont {{Turk}},
  \citenamefont {{Smith}}, \citenamefont {{Oishi}}, \citenamefont {{Skory}},
  \citenamefont {{Skillman}}, \citenamefont {{Abel}},\ and\ \citenamefont
  {{Norman}}}]{Turk2011}%
  \BibitemOpen
  \bibfield  {author} {\bibinfo {author} {\bibfnamefont {M.~J.}\ \bibnamefont
  {{Turk}}}, \bibinfo {author} {\bibfnamefont {B.~D.}\ \bibnamefont {{Smith}}},
  \bibinfo {author} {\bibfnamefont {J.~S.}\ \bibnamefont {{Oishi}}}, \bibinfo
  {author} {\bibfnamefont {S.}~\bibnamefont {{Skory}}}, \bibinfo {author}
  {\bibfnamefont {S.~W.}\ \bibnamefont {{Skillman}}}, \bibinfo {author}
  {\bibfnamefont {T.}~\bibnamefont {{Abel}}}, \ and\ \bibinfo {author}
  {\bibfnamefont {M.~L.}\ \bibnamefont {{Norman}}},\ }\href {\doibase
  10.1088/0067-0049/192/1/9} {\bibfield  {journal} {\bibinfo  {journal}
  {\apjs}\ }\textbf {\bibinfo {volume} {192}},\ \bibinfo {eid} {9} (\bibinfo
  {year} {2011})},\ \Eprint {http://arxiv.org/abs/1011.3514} {arXiv:1011.3514
  [astro-ph.IM]} \BibitemShut {NoStop}%
\end{thebibliography}
\end{document}